%\input bull-zag

%\font\cl=wncyrb10

%\font\mrm=cmr8
%\font\mbf=cmbx8
%\input bull-mac
%\pageno 37
%\def\ppp{37}
%\input mjcir
%\input mjyu
%\input astro
%\def\eps {\varepsilon }
%\def\sec {\buildrel ''\over.}
%\def\sev {\buildrel s\over.}

%\def\lhead{\mrm V. CELEBONOVIC}
%\def\rhead{\mrm THE ORIGIN OF ROTATION, DENSE MATTER PHYSICS AND 
%ALL THAT: A TRIBUTE TO PAVLE SAVIC}

{\ }

%\noindent {\bf Bull. Astron. Belgrade {\cl \char'175} 
%151 (l995), 37 -- 43} 
%\hfill\  {\bf UDC 523.2$-$327} 
%\rightline {Review paper}

{\ }

\vskip4.5cm
       
\centerline{\bf THE ORIGIN OF ROTATION, DENSE MATTER PHYSICS AND 
ALL THAT:}      
\vskip2mm
\centerline{\bf A TRIBUTE TO PAVLE SAVIC} 
\vskip.7cm
\centerline{\bf V. Celebonovic}
\vskip.7cm
\centerline{\it Institute of Physics, P.O.B. 57, 11001 Beograd, Yugoslavia} 
\vskip.7cm
\centerline{(Received: December 12, 1994)}
\vskip1cm

\parindent=2cm
\bf

\vbox{\narrower \noindent \hskip .5cm SUMMARY: This is a review of the main 
physical ideas and examples of applicability in astrophysics and pure physics 
of a semiclassical theory of dense matter proposed by Pavle Savic and Radivoje 
Kasanin in the early sixties. A hypothesis, advanced by Savic with the aim of 
solving the problem of the origin of rotation of celestical bodies,  will also 
be discussed. The paper is dedicated to the memory of Pavle Savic, who died 
recently. }

\parindent=1cm
\tenrm
\vskip1cm

%\dvekolone

\noindent {\bf 1.  INTRODUCTION}
\vskip.7cm

	   Dr.h.c. Pavle Savic, Emmeritus professor of physical 
chemistry at the University of Beograd, fou\-nder and first 
scientific director of the Institute of Nuclear Sciences in Vinca 
near Beograd, member and former President of the Serbian Academy 
of Sciences and Arts, died in Beograd on May 30, 1994. The purpose 
of this paper is to review some results of his scientific 
work, and, in an appendix, present the main details of his rich and 
interesting life. The title of the paper is an analogy with one 
of the most important books in modern theoretical physics 
(Streater and Wightman, 1964). 

Savic has been scientifically active in the fields of nuclear 
physics and dense matter physics and astrophysics. In nuclear 
physics,  he co-authored with Irene Joliot - Curie, some of the 
crucial experiments which have opened the way to the discovery of 
nuclear fission. His interest in dense-matter physics was 
motivated by the problem of the origin of rotation of celestical 
bodies. Savic formulated a hypothesis (Savic, 1961) which 
tried to contribute to the solution of this problem. After that, 
jointly with Professor Radivoje Kasanin, he developed a 
theory of the behaviour of materials under high pressure 
(Savic and Kasanin, 1962/65). This paper will present a 
review of the main ideas and results of their theory (called 
the SK theory). Readers interested in the "nuclear"activity of 
Pavle Savic should consult some of the existing excellent 
reviews of the subject (such as Seaborg, 1980; Savic, 1989).  
The paper by Savic is especially important,  because it gives 
a "first-hand" account of the conditions and style of work in the 
famous Institut du Radium in Paris, where he performed his 
research in nuclear physics.  

%\krajdvekolone

%\vfil\eject

%\dvekolone

\vskip.7cm

\noindent {\bf 2. THE SK THEORY}
\vskip.5cm

\noindent 2.1 {\it The origin of rotation}
\vskip.7cm

The starting point for the developement of this theory was a 
paper by Savic (Savic, 1961),  referred to in the following as 
S61, which had the aim of proposing an explanation of the origin 
of rotation of celestical bodies. As a result, there emerged the 
conclusion that rotation is tightly related to the internal 
structure, and that a correct theory of dense matter was needed.  

Considerations of S61 (and, of the whole SK theory) start from a 
low- temperature cloud con\-sis\-ting of an arbitrary number of 
chemical elements and their compounds. The life of such a cloud is 
governed by two physical processes:the mutual gravitational 
interaction of its constituting particles, and the loss of energy 
due to radiation. In modern terms, this is a description of a 
typical interstellar cloud.  

Due to the increase of pressure, two important processes occur in 
the interior of the cloud - the excitation and ionisation of  its 
constituting atoms and molecules. Qualitatively speaking, this is 
an expectable consequence of the perturbation of the electronic 
energy levels by the external pressure field. The physical 
possibility of this coupling can be pro\-ved  even in elementary 
quantum mechanics, on a solvable system such as a finite potential 
well. In quantum mechanics increasing the pressure to which a 
(macroscopic) system   is subjected leads to the expansion of the 
radial part of the electronic wave-function of the atoms and 
molecules that make up the system. It is interesting to note that 
the first fully quantum treatment of excitation and ionisation 
under pressure in a real solid body has been given only recently 
(Ma, Wang, Chen et al, 1988),  nearly three decades after the 
concept was used by SK.  

At a certain value of the pressure (which is, obviously, a 
function of the chemical composition of the primordial cloud) a 
phase transition occurs in the cloud - it passes into the state 
of a poly-component plasma (two-component in the simplest 
case). This plasma consists of a randomly moving electron gas and 
neutral and ionized atoms, molecules. General criteria for the 
existence of bound and ionized pro\-ton-electron           pairs in 
a plasma were recently discussed in detail 
(Girardeau, 1990; Lebowitz, Macris and Martin, 1992). It has been 
pro\-ved that such  an electron gas has a non-zero magnetic field   
(de Groot and Suttorp, 1972). Owing to a combination of high 
pressure and low-temperature, the magnetic moments of the ionized  
atoms and molecules become oriented 
 in paralel,  and the resulting torque starts the rotation of  the 
whole system. For a different approach to the problem of the 
origin of magnetic fields in the early solar system see, for 
example (Stepinski, 1991,  1992).  

The mechanism just described  may seem, at first sight, highly 
qualitative. However,  
 when elaborated in detail (Savic and Kasanin, 1962/65;   
Savic, 1981), it gives values of the magnetic fields and the 
allowed intervals of the speed of rotation of the Sun and the 
planets which are in good agreement with the observations. For 
example, the observed speed of solar rotation in the equatorial 
zone is $\omega$ = 2.9  10$^{-6}$ rad/s, while the physically allowed 
interval according to the SK theory is (1.2 $\le \omega \le$ 
44.7) 10$^{-6}$ 
rad/s.  The SK theory gives a value of 10-14 Gauss for the 
magnetic field of Jupiter, while the measured value is 14 Gauss (Savic, 1981).  

An obvious question emmerges at this point - why does the SK 
theory give the physically permitted interval and not a single 
value for the speed of rotation of a celestical body? This is due 
to the fact that the speed of rotation is calculated via the 
estimate of the work, done per atom, by the external pressure in 
expulsing the electrons from the energy levels in the atoms and 
molecules that make up the material. As the pressure is related to 
the density, and it has a lower and upper limit in every 
phase, the consequence is that the theory provides only the limits 
of the physically possible interval of the values of the speed of 
rotation.  

It would be an interesting problem in the\-ore\-ti\-cal plasma physics 
to reformulate this part of the SK theory, so as to narrow the 
limits of the interval, or even transform it in a single value. 
Another important question concerns lunar ro\-ta\-ti\-on. It is a known 
fact that the Moon rotates with a period equal to its orbital 
period. On the other hand, one often hears that, according to SK 
(Savic, 1972), the Moon should not rotate. The discrepancy  seems to 
be obvious and serious,  but it actually has a simple solution.  
Rotation of a body is, in the framework of SK,  a consequence of 
pressure ionization of atoms and molecules in its interior. The 
realizability of this process in a given celestical object 
depends on its mass and chemical composition. When they calculated 
a model of the internal structure of the Moon,          SK have determined 
the mean atomic mass of the chemical mixture that the Moon is 
made of (A = 71),  and the chemical elements which enter into 
its composition (B, Be, C, O, F, Ca Mg, Al, Si, Ti, V, Se, Y, Zr, Sb, Te, La 
and the rare earths). On the other hand,  it has been estimated in 
SK that the central pressure in the Moon is too low to ionize any 
of the chemical elements present there, which led to the 
theoretical conclusion that the Moon should not rotate. However,  
in their calculation, SK did not take into account the possibility 
that the Moon can contain molecules as well as pure elements, and 
that the ionization potentials of the molecules are often smaller 
than the ionization potentials of the pure atoms. This difference 
can be ascribed to the polarization of surrounding molecules 
and, in some cases, intermolecular band formation (Seki, 1989). For 
example, ionization potentials of Ca,  F and CaF are 6.113 
eV, 17.422 eV and 5.8 eV, respectively.  It follows that it is 
easier to ionize 
CaF by subjecting it to high pressure, than to achieve 
the same 
goal for      

%\krajdvekolone

%\vfil\eject

%\dvekolone

\noindent pure
Ca and F.  Reasoning in this way (that is, starting 
from the chemical elements proposed by SK for the composition of 
the Moon, forming molecules from them and then determining the 
ionization pressure), one could show that at least some of these 
molecules can be ionized in the interior of the Moon and that 
lunar rotation can be accomodated within the SK theory.  

\vskip.7cm

\noindent 2.2 {\it Materials under high pressure}
\vskip.7cm

The study of materials subjected to high pressure is important  in 
a variety of  situations in physics, astronomy and related 
sciences. These include the early Universe and phase transitions 
which occured in it (Satz, 1993, 1994),  heavy-ion collisions in 
accelerators      (Angert, Bourgarel, Brouzet et al, 1993;  
Geiger, 1994;  Schukraft, 1994),  the internal structure of  
stars,  planets and satellites (Zharkov and Trubitsyn, 1980;  
Celebonovic, 1990;  Lindblom,  1992;   Weber and 
Glendenning, 1993;   Sedrakian, Blaschke, Ropke and Schulz, 1994 
and numerous other publications), but also experiments performed 
in diamond-anvil cells         (such as Savic and Urosevic, 
1987), the synthesis of high-temperature superconductors  (Hiroi 
and Takano, 1994) and spectral-line broadening studies.  

Attempting the study of  a material subjected to high pressure 
and/or temperature is a difficult,  interesting but  {\bf 
solvable} 
problem. Experimental investigations encounter various kinds of 
technical problems (such as filling a diamond-anvil cell, or 
measuring the electrical conductivity of a specimen in  a cell).  
On the theoretical side, a specimen of a material is an example of 
a typical many-body system,  whose Hamiltonian can be expressed as

$$\rm H = \sum\limits_{i=1}^N - {\hbar^2\over{2m}} \nabla_i^2 + 
\sum\limits_{i=1}^N V (\vec{x_i}) + \sum\limits_{i,j=1}^N v 
(\vert \vec{x_i} - \vec{x_j}\vert ) \eqno{(1)}$$
	
All the symbols in Eq. (1) have their customary meanings. By 
using the standard approach of statistical mechanics, one should 
introduce some form of the interparticle potential $\rm v (\vert 
\vec{x_i} - \vec{x_j}\vert )$ in Eq. (1), calculate the 
partition function which is defined by ($\rm Z = tr \ exp (- \beta H)$), and from it 
the free energy and all the other required thermodynamic 
potentials (for example, Ruelle,  1969).   Instead of 
embarking on such a complicated calculation, Savic and 
Kasanin have founded their theory on a set of 6  
experimentally founded premisses supplemented by a microscopic 
selection rule.  

The mean interparticle distance $a$ is defined by the relation

$$\rm N_A (2a)^3 \rho = A \eqno{(2)}$$

\noindent where $\rm N_A$ denotes Avogadro' s number,  $\rho$ is 
the 
mass density and A is the mean atomic mass of the material. After thus 
introducing $a$, one can define the "accumulated" energy per 
electron as  

$$\rm E = e^2/a \eqno{(3)}$$

\noindent One might expect that a relation such as (3) should contain the 
ionic charge Z. It can be shown (Leung, 1984) that $a$, as defined 
in (2) is a multiple of the Wigner-Seitz radius,  which actually 
contains Z.  

The basic premisses of the SK theory are the following statements 
(Savic and Kasanin, 1962/65;   Celebonovic, 1989 
d): 

1. The density of a material is an increasing function of the 
pressure to which it is exposed.  

2. With increasing density, every material undergoes a sequence of 
phase transitions.  The phase ending at the critical point is 
denoted as the zeroth phase. For every phase, indexed by $i$, there 
exist two limiting values of the density, such that 

$$\eqalign{&{\rm \rho_i^0 \le \rho_i \le \rho_i^*}\cr
&{\rm (1/\alpha_i) \rho_i^* \le \rho_i \le \rho_i^* \ with \ 
\alpha_i > 1 \ .}\cr} \eqno{(4)}$$

3. The maximal densities of two successive phases 
are related by 

$$\rm \rho_{i+1}^* = 2 \rho_i^* \eqno{(5)}$$

\noindent which can be derived from the results of S61. It was shown in that 
paper that the mean densities of the planets can be fitted by the 
exppression 

$$\rm \rho_i =\rho_0 2^{\varphi_i} \eqno{(5a)}$$

\noindent where $\rho_0$ = 4/3 stands for the mean density of the Sun, and 
values of the exponent $\varphi_i$ i have to be chosen for each 
planet.  The density of a material is a consequence of its 
composition  and the equilibrium between the attractive and 
repulsive forces which exist in it. Values of $\varphi_i$ are a function 
of these equilibrium conditions and their changes under high 
external pressure. (5) follows from (5a) by assuming $\rm 
\varphi_{i+1} - \varphi_i = 1$.  

4. It is assumed that

$$\rm {E_i^*\over{E_i^0}}  =   {E_{i+1}^0\over{E_i^*}} 
\eqno{(6)}$$ 

Some form of a link between the accumulated energies in 
successive phases was needed in order to render the calculations 
possible, and (6) was accepted because of its simplicity. After 
some algebra, one gets that $\rm \alpha_i \alpha_{i+1} = 2$, and that 

$$\alpha_i = \left\{\eqalign{&{\rm 6/5 \ \ \ \ i = 1, 3, 
5,....}\cr
&{\rm 5/3 \ \ \ \ i = 2, 4, 6, ....}\cr}\right.\eqno{(6a)}$$

%\krajdvekolone

%\vfil\eject

%\dvekolone

5. The final density of the zeroth phase is

$$\rm \rho_0^* = {A\over{3\overline{V}}} \eqno{(7)}$$

\noindent which is approximately equal to the critical density in 
the van der Waals theory. $\rm \overline{V}$  denotes the molar 
volume of the material at T = 0 K;  in the  terminology of the 
van der Waals theory, $\rm \overline{V} = b$. 

6. Using assumption 3., it can be shown that

$$\rm {A\over{\overline{\rho}}} = {1\over 2} 
\left({A\over{\rho_2^0}} + {A\over{\rho_2^*}}\right) 
\eqno{(7a)}$$    

\noindent where $\rm \overline{\rho}$  denotes the density at the zero-point, defined as       
$\rm \overline{\rho} =  A/\overline{V}$.  

Starting from these premisses, the following set of relations can be derived:

$$\eqalign{&{\rm \rho_i^* = 2^i \rho_0^* \ \ \ \ \rho_i^0 = 
{\rho_i^*\over{\alpha_i}}}\cr
&{\rm V_0^* = 3\overline{V} \ \ V_i^0 = \alpha_i V_i^* 2^{-i} \ \ 
V_i^* = 2^{-i}V_0^*}\cr
&{\rm r_0^* = (15/4 N_A \times 10^{-23})^{1/3} \overline{V}^{-
1/3}}\cr
&{\rm r_i^* = 2^{-i/3} r_0^* \ \ \ \ r_i^0 = r_i^* \times 
\alpha_i^{1/3}}\cr}\eqno{(8)}$$ 

\noindent The SK theory has an inherent limitation - it can successfully 
treat only first order transitions. It is presumed in the theory 
that a discontinous change of volume $\rm V_i^* \to V_{i+1}^0$
occurs at the phase transitions pressure. Second order transitions 
can be
considered as a special case, for which $\rm V_i^* -  V_{i+1}^0 \to 0$

The value of the pressure at which a phase transition will occur 
in a given material can be calculated by considering the work 
done by the external pressure in compressing the material,  

$$\rm \Delta W = p_i^* (V_i^* - V_{i+1}^0) = p_i^* V_i^* 
(1-1/\alpha_i) \eqno{(9)}$$     

\noindent and equating it to the change of the accumulated energy

$$\rm \Delta W = \Delta E = N_A (E_{i+1}^0 - E_i^*) \eqno{(10)}$$

After some algebra,  manipulating equations (8) to (10), one arrives 
at the following expression for the maximal internal pressure in 
phase $i$ of a material having zero-point volume $\rm 
\overline{V}$: 

$$\rm p_i^* =  1. 8077 \beta_i (\overline{V})^{-4/3} 2^{4i/3} \ \  Mbar
\eqno{(11)}$$

\noindent where 

$$\rm  \beta_i = 3 {\alpha_i^{1/3} - 1\over{1 - 1/\alpha_i}} 
\eqno{(11a)}$$

The value of external pressure needed to induce a phase 
transition from phase $i$ to phase $i+1$ is, finally, given by 

$$\rm p_{tr} = p_i^* - p_i^0 = p_i^* \left(1 - 2^{-4/3} {\beta_{i-
1}\over{\beta_i}}\right)\eqno{(12)}$$ 

\noindent which can be reduced to

$$\rm p_{tr} = \left\{\eqalign{&{\rm 0. 5101 p_i^* \ \ \ \ i = 1,  
3,  5,... }\cr
&{\rm 0. 6785 p_i^* \ \ \ \ i = 2,  4,  6,... }\cr}\right.\eqno{(13)}$$

\noindent with  $\rm p_i^*$   determined by (11).  

Expression (13) represents a simple mathematical algorithm for 
the calculation of a sequence of  possible phase transition 
pressures in a given material. Those values of pressure at which a 
phase transition is physically possible according to SK are 
selected with the following criterion (Savic and 
Ka\-sa\-nin, 1962/65): 

$$\rm E_0^* + E_I = E_i^* \eqno{(14)}$$

\noindent $\rm E_I$ denotes the ionisation potential, $\rm E_0^*$ and $\rm 
E_i^*$ can be 
calculated by (3) using (8) and         the transformation $\rm a = r 
\times  10 ^{-8}$ cm.  

This procedure has recently been applied to a set of 19 materials 
of different chemical nature, with experimentally known phase 
transition points (Celebonovic, 1992 c). The unique criterion for 
the choice of the materials was the easy avaliability of 
experimental data. It was found that the agreement between the 
experimental and theoretical values of phase transition pressures 
is very encouraging - the relative differences vary between 
practically zero and 30 - 40\%. It may seem that the upper limit is 
very high, but it is actually of the same order of magnitude as 
the results of some quantum-mechanically founded calculations 
(Celebonovic, 1992c and references given there). The possible 
causes of the differences were analyzed in some detail. They 
include the contribution of various factors,  such as the exact 
form of the inter-particle potential in the material and the 
effects of  temperature. For details, see (Celebonovic 1992c).  

Apart from the calculation of the value of pha\-se transition 
pressure for a given material, another interesting (and partially 
open) problem in the SK theory concerns the establishement of 
the equation of state (EOS) of dense matter. In its original 
formulation,  this theory aims at proposing a representation of 
the so-called cold-compression curve, and 
the problem of the behaviour of dense matter with T $\ne$ 0 has 
been just touched in it. A form of the EOS in the $\rho - T$ plane 
has been proposed starting from the SK theory 
(Celebonovic, 1991b).  It was derived by comparing the 
expression for the accumulated energy unit volume with the result 
for the total energy per unit volume of a solid body, which is  
known  in solid state physics. As a test, this EOS was applied to 
the interior of the Earth,  and the results were in good 
agreement  with geophysical data. 

%\krajdvekolone

%\vfil\eject

\noindent {\bf Table 1:} The interior of the Earth
\vskip2mm

\settabs 5\columns

\+ depth (km) & 0  -  39  & 39  -  2900 & 2900  -  4980 & 4980  -  
6371&\cr

\+ $\rho_{max}$ (g/cm$^3$) & 3.0 & 6.0 & 12.0 & 19.74 &\cr

\+ $\rm P_{max}$ (Mbar) & 0.25 & 1.29 &   2.89 & 3.7 &\cr

\+ $\rm T_{max}$ (K) &  1300 & 2700 &   4100 & 7000  &\cr

\+ & & A = 26.56 & & &\cr

\vskip.5cm

%\dvekolone

\noindent However, proposing within SK a 
suitable form of the EOS of a thermo-mechanical system in the P - 
$\rho$ - T space of thermodynamical variables is currently an open 
problem.  

We have so far reviewed the applications of the SK theory in 
laboratory high-pressure work.  It can be used in 
astrophysics, where it gives the possibility of modelling the 
internal structure of celestical bodies. As input data it demands 
just the mass and radius of the object under study.  

Starting from these data, it provides the number of layers which 
exist in the interior of the object and their thickness, the 
distribution of pressure, density and temperature with depth under 
the surface, the strength of the magnetic field and the interval 
of the physically allowed values of the speed of rotation. It is 
very important that the SK theory gives the mean atomic mass of 
the chemical mixture  that the object under study is made of.  

The theory has so far been applied to the Sun, all the planets 
except Saturn and Pluto,  the Moon, the Galileian satellites, the 
satellites of Uranus, Neptune's satellite Triton and the asteroids 
1 Ceres and 10 Hygiea. The results are scattered in the 
literature, but, generally speaking the agreement with experiments 
and calculations performed by other methods is good. For 
example, it was calculated that the depth of the Moho layer in the 
Earth is 39 km, while the experimental value is 33 km. Tables 1 and 2 
contain data on the interiors of the Earth 
and the Moon calculated within the SK theory (Savic, 1972, 1981): 

\vskip.5cm

\noindent {\bf Table 2:} The internal structure of the Moon

\vskip2mm

\settabs 3\columns

\+ depth (km) & 0  -  338 & 338  -  1738 &\cr

\+ $\rm \rho_{max}$ &  3.32 & 6.64 &\cr

\+ P (Mbar) &   0.015 & 0.089 &\cr
  
\+ T (K) & 529 &  793 &\cr

\+ & A = 71 & &\cr

\vskip.5cm

The values of  T in table 2 were calculated by using eq. (7) of 
(Celebonovic, 1991b) and assuming Z = 1. 

Parameters of the interiors of various Solar System bodies 
calculated within the SK theory can not be compared to any direct 
experiments, but only to the consequences of these results on the 
visible layers of these objects. On the other hand, observationally 
verifiable conclusions can be drawn from an analysis of the mean 
atomic masses, $<$ A $>$,  of the chemical mixtures making up objects 
in the planetary system. A review of all the currently avaliable 
values of $<$A$>$  determined within the SK theory is presented in 
the following Table. The symbols {\it Ji} ({\it i} = 1,.. 4) and 
{\it Uk} ({\it k} 
=1,... 5) denote the satellites of  Jupiter and Uranus. 

\vskip.5cm

\noindent {\bf Table 3:} the chemical composition of some Solar System bodies
\vskip2mm

\settabs 4\columns

\+  object & $<$A$>$ &   satellite &    $<$A$>$  &\cr
\vskip3mm

\+  Sun &   1. 4 & Moon &         71                &\cr
\+  Mercury  & 113 &          J1            &  70&\cr
\+  Venus &        28. 12  &   J2 &              71&\cr
\+    Earth        &  26. 56 &    J3 &             18&\cr
\+  Mars       &   69        &   J4    &          19&\cr
\+  1 Ceres     & 96   &        U1 &            38&\cr
\+  Jupiter        &  1. 55    &  U2   &          43&\cr
\+ Saturn          &  / &          U3   &          44&\cr
\+ Uranus     &      6. 5   &    U4         &    32&\cr
\+ Neptune     &    7. 26  &    U5 &            32&\cr
\+ Pluto        &       /       &   Triton  &       67&\cr

\vskip.5cm                                                     

It can easily be seen from the preceeding Table that the 
planetary system is chemically inhomogenous;  the well known 
division on the terrestrial and jovian planets is  clearly 
visible.  The asteroid 1 Ceres,  whose orbit is currently between 
those of  Mars and Jupiter is,   by its chemical composition,  
similar to Mercury (Celebonovic,  1988b).  Physically,  this 
could imply that "once upon a time"  it originated close to 
Mercury,  but that their orbits later diverged (chaos ??).  Using 
the value of $<$A$>$ calculated for  1 Ceres within SK,  the mass of 
the asteroid  10 Hygiea  was calculated,  and the result is in 
excellent agreement with the experimental value known in 
celestical mechanics.   

A similar case exists in the Neptune - 
Triton system;  their respective values of  $<$A$>$ differ by almost 
an entire order of magnitude (Celebonovic,  1986).  This 
can be interpreted as a consequence of the fact that Triton is a 
captured body.  A similar conclusion follows from the analyses of 
its motion.   

Gradients in $<$A$>$  visible in the jovian and uranian 
satellite systems have been interpreted as a result of various 
transport processes in the respective circum-planetary accretion 
disks (Celebonovic,  1987;   1989).   

The problem of the 
origin of the Earth's satellite Moon has been considered within the SK theory 
(Savic and Teleki,  1986).  Indications were presented in 
favour of the common origin of the two bodies.   

%\krajdvekolone

%\vfil\eject

%\dvekolone

\vskip.7cm

\noindent {\it Instead of a conclusion}
\vskip.7cm

At the end of this  review of the theory of dense matter proposed 
by Pavle Savic and Radivoje Kasanin,  a few general 
comments are in order.   

Distinct advantages of their theory are 
its phy\-si\-cal and mathematical simplicity,   which were made 
possible by a careful choice of the starting assumptions.  For 
those acustomed to the language of modern theoretical physics,  
the SK  theory can be described as a mean field type theory based 
on the Coulomb interaction,  and supplemented by a microscopic 
selection rule.   The simplicity of the theory necessarily induces 
a certain amount of coarsness in its results.  One such example is 
the fact that the theory does not give a unique value for the 
speed of rotation of the celestical object under study,  but,  
instead an interval of physically allowed values.  Some work 
aimed at refining the theory is already under way.  It is hoped 
that this review will motivate (at least some of) the readers to 
get acquainted  with the details of the SK theory and try to 
perform some calculations of their own.   

\vskip.7cm

\noindent {\bf APPENDIX:} \ some biographical notes on Pavle Savic
\vskip.7cm

Pavle Savic was born on January 10,  1909 in Thessaloniki (Greece),  
where his  father was an employee of the Customs service of the 
Kingdom of Serbia.  At the age of 23,   he majored in physical 
chemistry at the University of Beograd,  where he became a teaching 
assistant.  His first scientifc publication,  written in 1931.  with 
his professor Dragoljub Jovanovic,  dealt with the methods of  
measurements of $\gamma$ rays.  In 1935 he was awarded a 
French Governement schollarship for a 6 months visit to the 
Institut du Radium in Paris,   one of the world's best centers for 
research in nuclear physics at that time.  Instead of 6 months,  he 
stayed for 4 years and contributed,  with Irene Joliot-Curie,  to 
some of the historic experiments which have led to the discovery 
of nuclear fission.   He was "asked to leave French territory" 
because of the sensitive pre-war political situation in 
France.  When war began in Yugoslavia,  he joined the partizan 
movement and was assigned various important positions in the General 
Headquarters.   

After the war,  in 1945.,  he became a professor of 
physical chemistry at the University of Beograd,  and a year later 
was elected a corresponding member of the Serbian Academy of 
Sciences (as it was then called).  As an important act,  Savic 
initiated the founding of the Institute for the study of the 
structure of matter in Vinca near Belgrade.  He assured the 
scientific direction of the Institute until  1960.,  and remained 
at the University until 1966.  For 10  years,  starting in 1971.,  he 
was President of the Serbian Academy of Sciences and Arts,  and 
managed,  during  that time,  to completely modernize its 
functioning.   Prof.  Savic gave the impulse to the foundation,  in 
the Institute of Physics,  of a high-pressure physics 
laboratory.  He remained scientifically active (within the 
limitations imposed by his age and health) literally until the 
end of his life.  His final publication (Savic and 
Celebonovic,  1994),  appeared a few months after his death.   

Pavle 
Savic was a scientist who managed to combine strict 
knowledge,  bold original ideas and an intuitive "sense" of 
nature,  an excellent teacher and a trusted friend.  He will be 
remembered and missed by all those who knew him.   

\vskip.7cm

\noindent {\it Acknowledgment} -- Thie preparation of this 
review has been financed by the Ministry of Science and 
Technology of  the Republic of Serbia.  

\vskip.7cm 

\noindent REFERENCES
\vskip.7cm 

\item{}\kern-\parindent{Angert, N., 
Bourgarel, M. 
P., Brouzet, E., Cappi, R., Dekkers, D., Evans, J. et al: 1993, {\it CERN 
Heavy-ion facility design report}, CERN 93-01.} 

\item{}\kern-\parindent{Celebonovic, V.: 1986,  {\it Earth, Moon and 
Planets}, {\bf 34}, 59. }

\item{}\kern-\parindent{Celebonovic, V.: 1987, presented at the II 
Workshop " Astrophysics in Yugoslavia",  
 (Astronomical 
Observatory, Belgrade, September 8.-10.) and 
published in the book of abstracts, p. 41.} 

\item{}\kern-\parindent{Celebonovic, V.: 1988b, {\it Earth, Moon and 
Planets}, {\bf 42}, 297. }

\item{}\kern-\parindent{Celebonovic, V.: 1989d, {\it Earth, Moon and 
Planets}, {\bf 45},  291. }

\item{}\kern-\parindent{Celebonovic, V.: 1989, {\it Bull. Obs. Astron. 
Belgrade}, {\bf 140}, 47. }

\item{}\kern-\parindent{Celebonovic, V.: 1990, {\it High Pressure 
Research}, {\bf 5}, 693. }

\item{}\kern-\parindent{Celebonovic, V.: 1991b, {\it Earth, Moon and 
Planets}, {\bf 54}, 145. }

\item{}\kern-\parindent{Celebonovic, V.: 1992c, {\it Earth, Moon and 
Planets}, {\bf 58}, 203. }

\item{}\kern-\parindent{de Groot, S. R. and Suttorp, L. G.: 
1972, {\it Foundations of Electrodynamics}, North Holland Publ. Comp., 
Amsterdam.} 

\item{}\kern-\parindent{Geiger, K.: 1994,  {\it preprint CERN-TH.} 7429/94. }

\item{}\kern-\parindent{Girardeau, M. D.: 1990, {\it Phys. 
Rev.}, {\bf A41}, 6935. }

\item{}\kern-\parindent{Hiroi, Z. and Takano, M.: 1994,  
{\it preprint, to appear in Physica C} (proceedings of 
the 
 Grenoble M2 S - HTSC IV Conference).}

\item{}\kern-\parindent{Lebowitz, J. L, Macris, N. and Martin, P. 
A.: 1992, {\it J. Stat. Phys.}, {\bf 67}, 909. }

\item{}\kern-\parindent{Leung, Y. C.: 1984, {\it Physics of 
Dense Matter}, Science Press/ World Scientific, Beijing and Singapore. }

\item{}\kern-\parindent{Lindblom, L.: 1992, {\it Astrophys. J.}, {\bf 398}, 
569. }

%\krajdvekolone

%\vfil\eject

%\dvekolone

\item{}\kern-\parindent{Ma, D., Wang, Y., Chen, J. and 
Zhang, Y.: 1988, {\it J. Phys.}, {\bf C21}, 3585. }

\item{}\kern-\parindent{Ruelle, D.: 1969, {\it Statistical 
Mechanics}, W. A. Benjamin Inc., London. }

\item{}\kern-\parindent{Satz, H.: 1993, {\it preprint CERN - 
TH.} 7064/93. }

\item{}\kern-\parindent{Satz, H.: 1994,  {\it preprint CERN - 
TH.} 7410/94. }

\item{}\kern-\parindent{Savic, P.: 1961, {\it Bull. de la classe 
des Sci. Math. et natur. de l'Acad. Serbe des 
Sci. et des 
 Arts}, {\bf XXVI}, 107.}

\item{}\kern-\parindent{Savic, P. and 
Kasanin, R.: 1962/65, {\it The Behaviour of Materials 
Under High Pressure }
 I-IV, Ed. by SANU, Beograd. }

\item{}\kern-\parindent{Savic, P.: 1972, {\it C.R. des seances de la 
Soc. Serbe de Geologie pour 1968.- 1970,} p. 279. }

\item{}\kern-\parindent{Savic, P.: 1981, {\it Adv. in Space Res.}, {\bf 1}, 
131. }

\item{}\kern-\parindent{Savic, P. and Teleki, G.: 1986, {\it Earth, Moon 
and Planets}, {\bf 36}, 139. }

\item{}\kern-\parindent{Savic, P. and 
Urosevic, V.: 1987, {\it Chem. Phys. Lett.}, {\bf 135}, 393. }

\item{}\kern-\parindent{Savic, P.: 1989, in: {\it 50 years with nuclear 
fission}, ed. by J. W. Behrens and A. D. Carlson,  
 p. 20, Ed. by the American Nuclear Society 
Inc., La Grange Park, IL 60525 USA.} 

\item{}\kern-\parindent{Savic, P. and Celebonovic, V.: 1994,  
in: {\it Proc. of the 1993. Joint AIRAPT/APS Con. on 
High Pressure Sci. and Technol.},  AIP conference 
proc., {\bf 309}, p. 53, AIP Press, New York.} 

\item{}\kern-\parindent{Schukraft, J.: 1994, {\it preprint CERN} - PPE/94-139.}

\item{}\kern-\parindent{Seaborg, G. T: 1980, in: {\it Collection of 
papers devoted to Pavle Savic on the 
 occasion of his seventieth birthday}, ed. by M. 
Garasanin, p. 39., Ed by SANU, Beograd.} 

\item{}\kern-\parindent{Sedrakian, A. D., Blaschke, D., Ropke, D. 
and  Schulz, H.: 1994, {\it Phys. Lett.}, {\bf B338}, 111. }

\item{}\kern-\parindent{Seki, K.: 1989,  {\it Mol. Cryst. Liq. 
Cryst.}, {\bf 171},  255. }

\item{}\kern-\parindent{Stepinski, T. F.: 1991, {\it Publ. Astron. Soc. 
Pacific}, {\bf 103}, 777. }

\item{}\kern-\parindent{Stepinski, T. F. 1992, {\it Icarus}, {\bf 97}, 130.} 

\item{}\kern-\parindent{Streater, R.F. and Wightman, A. S.: 1964, {\it PCT, spin 
and statistics and all that}, Benjamin, New York.} 

\item{}\kern-\parindent{Weber, F. and Glendenning, N. K.: 
1993, {\it Lawrence Berkeley Lab.}, preprint LBL- 34783. }

\item{}\kern-\parindent{Zharkov, V. N. and Trubitsyn, V. P.: 
1980, {\it Physics of Planetary Interiors} (in Russian) 
 Nauka Publ. House, Moscow.}

%\krajdvekolone

{\ }

\vskip2cm

%\rrm

\bye